
\magnification=1200
\overfullrule=0pt
\baselineskip=20pt
\parskip=0pt
\def\dag{\dagger}
\def\del{\partial}

\def\d{\delta}     
\def\e{\epsilon}   
\def\z{\zeta}

\def\m{\mu}

\def\r{\rho}       
\def\s{\sigma}

\def\f{\phi}       
       
\def\y{\psi}       
\def\w{\omega}     
\def\br{\langle}
\def\ke{\rangle}
\def\ve{\vert}

{\settabs 5 \columns
\+&&&&\cr}
\bigskip
\centerline{\bf EFFECTIVE INTERACTIONS OF PLANAR FERMIONS }
\centerline{\bf IN A STRONG MAGNETIC FIELD-THE EFFECT OF LANDAU LEVEL MIXING}
\bigskip\bigskip
\centerline{Rashmi Ray$^1$ and Gil Gat$^2$}
\bigskip

\centerline{ Department of Physics }
\centerline{ University of Maryland at College Park }
\centerline{ College Park, MD 20742-4111, USA }
\bigskip
\centerline{\bf Abstract}
\bigskip
We obtain expressions for the current operator in the lowest Landau level
(L.L.L.) field theory, where higher Landau level mixing due to various
external and interparticle interactions is sytematically taken into account.
We consider the current operators in the presence of electromagnetic
interactions, both Coulomb and time-dependent,
as well as local four-fermi interactions. The importance of Landau level
mixing for long-range interactions is especially emphasized. We also
calculate the edge-current for a finite sample.

\vfill
\noindent{$^1$E-mail address: rray@delphi.umd.edu.}

\noindent{$^2$E-mail address: ggat@delphi.umd.edu.}
\vfill\eject
\centerline{\bf I. Introduction}
\bigskip
The many-body problem of charged planar fermions is a fascinating problem
in its own right. It is encountered, for example, in the quantum Hall
phenomena [1] as well as in the theory of $c=1$ strings, where the connection
is not an obvious one [2].
A strong magnetic field creates a large gap between successive Landau levels
of the single-particle spectrum. Specifically, the gap between the lowest
Landau level (L.L.L.) and higher levels is of the order of the cyclotron
frequency, which is high for a large magnetic field. This naturally indicates
that a second quantized field theory of the lowest Landau level fermions
is the most convenient tool for handling the problem in a strong magnetic field
{}.

Since the seminal work of Girvin \& Jach, [3], a lot of progress has been made
in this direction [1],[5]. As the degenerate states of the L.L.L. do not form
a complete set, the L.L.L. field operators do not satisfy the standard
fermionic anticommutation relation. In fact, the field operators at equal time
but at different spatial points are related. This feature makes the formulation
of a second quantized theory a challenging task. Specifically, as has been
demonstrated by Stone [1], the current operator for the L.L.L. fermions is
not the operator one would expect in a more standard fermionic field
theory. Recently, the appropriate current operators have been obtained,
for the cases of an external local interaction and for a Coulomb interaction
between the fermions [4],[5]. However, in these works, the effect of mixing
with higher L.L. through the perturbing potentials has not been addressed. As
indicated in [6], the effect of such mixing could be significant in some
physical situations.
In the sequel, we have constructed expressions for the current operator
for the L.L.L. fermions in the presence of a variety of interactions,
where Landau level mixing has been systematically accounted for.
Our results are an explicit realization of the idea schematically discussed
in [6].

The organization of this paper is as follows: in section II we set
up the basic notation, in III we obtain an effective action for the L.L.L
fermions and in the subsequent sections we utilize this action to obtain
the current operators in the presence of Coulomb (IV), electromagnetic (V)
and local four-Fermi (VI) interaction. In VII we discuss the edge current
in the case of a finite sample.
\bigskip
\centerline{\bf II.  Notations and Conventions}
\bigskip
The kinematics of electrons in the L.L.L. has already been discussed [3].
However, to make the discussion reasonably self-contained, we introduce our
own notation [7] in the following.

The Lagrangian for planar electrons in a magnetic
field normal to the plane is
$$ \int d \vec x\ \y^\dag (\vec x, t) \biggl(i \del_t -h_0\biggr)
\y (\vec x, t). \eqno(2.1)$$
The free single-particle Hamiltonian is given by
$$ h_0 = {1\over{2m}}[({\hat \pi}^x)^2 + ({\hat \pi}^y)^2 ], \eqno(2.2)$$
where ${\hat \pi}^i \equiv \hat p_i-A^i(\hat x,\hat  y)\ ;\ \vec A \equiv
{B\over2}(y,-x)\ $ and the constant magnetic field is in the negative
$\hat z$ direction.
We define $$ \hat \pi \equiv {1\over{\sqrt{2B}}}({\hat \pi}^x -i{\hat \pi}^y)
\eqno(2.3).$$
$\hat \pi, \hat \pi^\dag $ are dimensionless. Their commutator is
$$[\hat \pi , \hat \pi^\dag ]=1 \eqno(2.4)$$
In this notation,
$$h_0=\w \hat \pi^\dag \hat \pi \eqno(2.5),$$
where we have dropped the zero point fluctuation and $\w = {B\over m}$,
the cyclotron frequency.
Further, define the guiding-centre coordinate operators as
$$\hat X \equiv \hat x -{1\over B}\hat \pi^y\ ;\ \hat Y \equiv \hat y + {1\over
B}\hat \pi^x \eqno(2.6).$$
We also define their holomorphic combinations
$$\hat a \equiv \sqrt {{B\over2}}(\hat X +i \hat Y)\ ;\ \hat a^\dag \equiv
\sqrt {{B\over2}}(\hat X -i \hat Y)\eqno(2.7)$$
for which the commutators are:
$$[\hat X,\hat Y]={i\over B},\  [\hat a,\hat a^\dag]=1 \eqno(2.8).$$
Also,
$$[\hat a,\hat \pi]=[\hat a^\dag,\hat \pi]=[\hat a,\hat \pi^\dag]
=[\hat a^\dag,\hat \pi^\dag]=0\eqno(2.9).$$
Thus there are two pairs of independent canonical operators defined on
this Hilbert space. Of them, $\hat a$ and $ \hat a^\dag $ commute with $h_0$
and thus characterize the degeneracy of the landau levels.
Let us further define
$$\hat z \equiv \sqrt {{B\over2}}(\hat x+i\hat y)=\hat a-i\hat \pi^\dag \
;\ \hat {\bar z}=\hat z^\dag \ ;\ [\hat z,\hat {\bar z}]=0.\eqno(2.10)$$
We specify the coordinates of the Hilbert space by $\ve \vec x \ke$ or
$\ve n \ke \ve \z \ke ,$ where
$$\eqalignno{\hat \pi^\dag \hat \pi \ve n \ke &=n\ve n \ke \cr
\hat a \ve \z \ke &=\z \ve \z \ke .&(2.11)\cr } $$
Here $\ve \z \ke \equiv e^{\z \hat a^\dag }\ve 0 \ke,\ \hat a \ve 0 \ke
=0.$
These two are related by
$$\br \vec x \ve (\ve n \ke \ve \z \ke)={\sqrt{B\over{2\pi}}}
i^n {{(z-\z )^n}\over{\sqrt{n!}}}e^{-{1\over2}\ve z \ve^2 + \bar z \z}
\eqno (2.12),$$
where $z \equiv {\sqrt {B\over 2}}(x+iy).$ From 2.12 it is clear that for n=0
the $\z$ enables one to make the transition to the coordinate most easily.

We represent the second-quantized Schr\"odinger operator by $\ve \y \ke$,
where $\hat \y(\vec x, t)\equiv \br \vec x,t \ve \y \ke $ is the
corresponding field operator.
We can write
$$\ve \y \ke =\sum_{\rm n} \ve \y_n \ke \eqno(2.13), $$
where $\ve \y_n \ke$ is the Schr\"odinger operator for the nth Landau
level.
It follows that $\hat \pi \ve \y_0 \ke =0,$ which means that
$$\br \vec x \ve \hat \pi \ve \y_0 \ke = -i(\del_z + {1\over2}\bar z)
\hat \y_0 (\vec x, t)=0. \eqno(2.14)$$
This is the L.L.L. condition satisfied by the L.L.L. Schr\"odinger
field
$$\hat \y_0 (\vec x,t)={\sqrt{B\over{2\pi }}}e^{-{1\over2}\ve z \ve^2
}\sum_{\rm l=0}^\infty {{{\bar z}^l}\over{\sqrt{l!}}}\hat c_l(t),
\eqno(2.15)$$
where $\{ \hat c_l, {\hat c_{l^\prime}}^\dag \} =\d_{ l,l^\prime }.$
Let us further define
$$\hat X \ve x \ke  =x \ve x \ke  ,\ \hat Y \ve y \ke =y \ve y \ke ,$$
where
$$\br x \ve y \ke = {\sqrt {B\over{2\pi }}}e^{iBxy}\eqno(2.16)$$
and
$$\eqalignno{\br x \ve \z \ke &=({B\over{\pi}})^{{1\over4}}e^{-{1\over2}
(Bx^2 + z^2)+{\sqrt{2B}}x \z }\cr
\br y \ve \z \ke &=({B\over{\pi}})^{{1\over4}}e^{-{1\over2}
(Bx^2 - z^2)-i{\sqrt{2B}}y \z }.&(2.17)\cr}$$
It is sometimes convenient to choose an angular momentum basis for the single
-particle states, given by
$$\hat a^\dag \hat a \ve l \ke = l\ve l \ke ,\ \br l \ve l^\prime \ke
=\d_{ l,l^\prime},\ \br l \ve \z \ke = {{{\z}^l}\over{\sqrt{l!}}}.
\eqno (2.18)$$
$\hat c_l $ in (2.15) above destroys a L.L.L. electron with angular
momentum $l$.

Having established the notation, let us introduce an interaction
$V(\vec x,t)$ into the single-particle problem.
Thus,
$$h=h_0 + V. \eqno(2.19)$$
Now, a general scalar potential $V$ can be written as
$$\eqalignno{V(\vec x,t)&= V(\hat z, \hat z^\dag ,t)\cr
&= \sum_{\rm m,n}{1\over{m! n!}}(-i)^n (i)^m (\hat \pi^\dag )^n
(\hat \pi )^m \ddagger \del_z^n \del_{\bar z}^m V(z, \bar z,t)\ve \hat
z =\hat a^\dag , \hat {\bar z}=\hat a \ddagger .&(2.20)\cr }$$
The ordering adopted for $\hat \pi , \hat \pi^\dag $ automatically forces
us to anti-normal order $\hat a,\hat a^\dag $. This is indicated by
$\ddagger \ \ddagger .$
We thus write
$$V(\hat{\vec x},t)=\sum_{\rm m,n=0}^\infty (\hat \pi^\dag )^m(\hat \pi)^n
v_{\rm m,n}\eqno(2.21)$$
Where,
$$\eqalignno{v_{00}&=\ddagger V(\hat a, \hat a^\dag ,t)\ddagger \cr
v_{10}&=-i\ddagger \del_z V \ddagger = v_{01}^\dag \cr
v_{11}&=\ddagger \del_z \del_{\bar z}V\ddagger \cr
v_{20}&={{(-i)^2}\over{2!}}\ddagger \del_z^2 V \ddagger = v_{02}^\dag
.&(2.22)\cr }$$

\centerline{\bf III Effective Action for the L.L.L.}
\bigskip
In the notation established above, the action is written as
$$-S=\br \y \ve \hat p_t + \hat h \ve \y \ke . \eqno(3.1)$$
Here, $\br t \ve \hat p_t \equiv -i \del_t \br t \ve .$
More explicitly
$$\eqalignno{-S &=\br \y_0 \ve \hat p_t + \hat V \ve \y_0 \ke +
\sum_{\rm n \neq 0}
[\br \y_0 \ve \hat V \ve \y_{\rm n}\ke + \br \y_{\rm n}\ve \hat V \ve
\y_0 \ke ]\cr &+ \sum_{\rm n,n^\prime \neq 0}\br \y_{\rm n}\ve \hat p_t
+ \hat h \ve \y_{\rm n^\prime }\ke ,&(3.2)\cr }$$
where we have used $\hat h_0 \ve \y_0 \ke =0.$
The L.L.L. truncation corresponds to keeping only the first term
on the r.h.s. of (3.2). However, in order to incorporate the effect
of Landau level mixing, we integrate the $ \rm n \neq 0 $ modes
out instead of simply dropping them from the problem. This leads to
an effective action
$$-S_{\rm eff}=\br \y_0 \ve \hat p_t + \hat V \ve \y_0 \ke -
\br \y_0 \ve \hat V \lq\lq {1\over{\hat p_t + \hat h_0 + \hat V}}"\hat V
\ve \y_0 \ke \eqno(3.3),$$
where $\lq\lq  $" indicates that $\rm n=0$ does not enter into the sum over
intermediate states.

At this point, we choose to work with potentials that are slowly varying
in space and time. Further, the potentials are taken to be small in
magnitude compared to the cylcotron frequency.

This naturally leads to the expansion
$$\lq\lq {1\over{\hat p_t +\hat h_0 +\hat V }}"=\lq\lq {1\over{h_0}}"
-\lq\lq {1\over{h_0}}"
(\hat p_t + \hat V )\lq\lq {1\over{h_0}}"+\dots \ .\eqno(3.4)$$
In what follows we drop the subscript 0 and the karets over the operators
as they should be clear from the context. We also drop the $\lq\lq $" and the
absence of $\rm n=0$ is to be tacitly understood.
Thus,
$$-S_{\rm eff}=\br \y \ve(1+V{1\over{{h_0}^2}}V)
p_t \ve \y \ke + H_{\rm eff}^{(0)} + H_{\rm eff}^{(1)} + H_{\rm eff}^{(2)} +
\dots . \eqno(3.5)$$
Here,
$$\eqalignno{H_{\rm eff}^{(0)} &\equiv \br \y \ve V \ve \y \ke \cr
H_{\rm eff}^{(1)} &\equiv -\br \y \ve V {1\over{h_0}}V \ve \y \ke \cr
H_{\rm eff}^{(2)} &\equiv -i\br \y \ve V {1\over{{h_0}^2}}\del_t V \ve \y \ke
\cr H_{\rm eff}^{(3)} &\equiv \br \y \ve V {1\over{h_0}}V{1\over{h_0}} V
\ve \y \ke .&(3.6)\cr }$$

At this point we see that if we keep the magnetic field fixed in value
and work with a large cyclotron frequency, we have to let $m \rightarrow
0.$ Then  $m / \sqrt{B}$ is a natural expansion parameter. We obtain $S_{\rm
eff}$
to $ O(m / \sqrt{B})$, where $H_{\rm eff}^{(0)}$, which corresponds to a L.L.L.
truncation is of O(1). One observes that to  order $O(m/\sqrt{B})$
the symplectic structure of the effective action retains its original form
and the fermion field need not be renormalized to regain its original
symplectic structure.

Thus, the effective action to $O(m / \sqrt{B})$ is
$$S_{\rm eff}=-\br \y \ve p_t \ve \y \ke - H_{\rm eff}^{(0)} -H_{\rm eff}
^{(1)}\eqno(3.8).$$
Explicitly,
$$\eqalignno{S_{\rm eff}&=-\int dt\ \int d \vec x \bar \y(\vec x,t)
i\del_t \y(\vec x,t)-\int dt\ \int d^2z \r(z,\bar z,t)\bigl[
V(z,\bar z,t)\cr
&-{1\over{\w }}\sum_{\rm n=1}^\infty {1\over{{n!}n}}\sum_{\rm m=0}^\infty
{{(-1)^m}\over{m!}}(\del_{\bar z}^{n+m}V)(\del_z^{n+m}V)\bigr] .&(3.9)\cr }$$
Thus $H_{eff}$ can be cast in the form
$$H_{\rm eff}\equiv \int d\vec x\ \r(\vec x,t)U(\vec x,t)\eqno(3.10)$$
where
$$U(\vec x,t)=V(z,\bar z,t)-{1\over{\w }}\sum_{\rm n=1}^\infty
\sum_{\rm m=o}^\infty {{(-1)^m}\over{m! n! n}}(\del_{\bar z}^{n+m}V)
(\del_z^{n+m}V)\eqno(3.11).$$
\bigskip
\bigskip
\centerline{\bf IV Current Operators and Landau Level Mixing}
\bigskip
In this section we utilize $S_{eff}$ to derive expressions for the current
opertors beyond leading order. To this end we use the equation of continuity
in cojunction with the Heisenberg equation of motion.

$$\eqalignno{\del_t \r(\vec x,t)&=-i[\r(\vec x,t),H_{\rm eff}]\cr
&=-i\int d\vec x^\prime \ [\r(\vec x,t),\r(\vec x^\prime ,t)]U(\vec x^\prime ,t
)\cr
&=-i\sum_{\rm n=1}^\infty {1\over{n!}}[\del_{\bar z}^n \{ \r(\vec x,t)
\del_z^n U(\vec x,t)\}-\del_z^n \{ \r(\vec x,t)
\del_{\bar z}^n U(\vec x,t)\} &(4.2)\cr }$$
where we have used
$$[\r(\vec x,t),\r(\vec x^\prime ,t)]=\sum_{\rm n=1}^\infty
{1\over{n!}}[\del_{\bar z}^n \{ \r(\vec x,t)\del_z^n \d(\vec x-\vec x^\prime )
\}-\del_z^n \{\r(\vec x,t)\del_{\bar z}^n \d(\vec x-\vec x^\prime ) \}]
\eqno(4.1).$$
which has been derived in [2],[5]. Using
$$\del_t \r(\vec x,t)=-\del_{\bar z}J(\vec x,t)-\del_z \bar J(\vec x,t)
\eqno(4.3)$$
and combining (4.1) and (4.2), we get
$$J(\vec x,t)=i\sum_{\rm p=0}^\infty \sum_{\rm n=1}^\infty \sum_{\rm q=1}
^\infty {{(-1)^p}\over{n! q! p!}}\del_{\bar z}^{n-1}[\r(\vec x,t)
\del_z^n \{ V(\vec x,t)-{1\over{\w q}}(\del_{\bar z}^{q+p}V)
(\del_z^{q+p}V)\}]\eqno(4.3),$$
to $O(m/ \sqrt{B})$.
The first term, obtained solely from $H_{\rm eff}^{(0)}$, does not contain
any information about Landau level mixing, has been obtained earlier [5].
It is the second term which in principle contains the effect of all higher
Landau levels, that is interesting for the study of mixing. For simplicity,
we  assume the potential to be slowly varying in space-time and
drop terms containing more than two derivatives.

Thus, to leading order,
$$J(\vec x,t)\simeq i\r(\vec x,t)\del_z [V(\vec x,t)-{1\over{\w }}
(\del_{\bar z}V)(\del_z V)]\eqno(4.4)$$
 From these we easily obtain
$$J^i(\vec x,t)={1\over B}\r(\vec x,t)\e^{0ij}\del_j [V-{m\over{2B^2}}
(\vec \nabla V)^2]\eqno(4.5).$$
\bigskip
\centerline{\bf V. Coulomb Interaction between the Electrons}
\bigskip
Let us consider a potential of the form
$$V(\vec x,t)=V_0(\vec x,t)+a_0(\vec x,t)\eqno(5.1).$$
Here, $a_0$ is a potential leading to an electric field. We imagine
that there is no perturbative magnetic fields acting on the electrons.
Thus we set $a_{i}$ to zero. The potential $a_0$ has an associated
Maxwell term given by ${1\over 2}f_{i0}f^{i0}.$ At this point
one should bear in mind that even though the problem is planar, the
electrons move in a two-dimensional subspace of a three-dimensional
space. Consequently, the Coulomb potential between charges is
of a ${1\over r}$ form rather than a log r form. To implement this
analytically, we extend the problem to three spatial dimensions.
Namely,
$$\eqalignno{\r(x,y,t)&\rightarrow \r(x,y,z,t) \equiv \r(x,y,t) \d(z) \cr
a_0(x,y,t)&\rightarrow a_0(x,y,z,t) \equiv a_0(x,y,t)\d(z) .&(5.2)\cr }$$
The equation of  motion for $a_0$ as obtained from the Hamiltonian,
including the Maxwell term is given by
$${\nabla }^2a_0 =-\r -{m\over{{B}^2}}[\r {\nabla }^2a_0 + \nabla
\r \cdot \nabla a_0] \eqno(5.3)$$
Here we have set $V_0$ to zero for simplicity.
This yields, to O(1),
$$a_0(\vec x,t)={1\over{4\pi }}\int d\vec {x^\prime }\ {\r(\vec {x^\prime },t)
\over{\ve \vec x-\vec {x^\prime  }\ve }}\eqno(5.4).$$
Putting this back into (5.3), we obtain, to $O(m/\sqrt{B})$,
$${\nabla }^2a_0=-\r_{\rm eff}\eqno(5.5),$$
where
$$\r_{\rm eff}\equiv \r - {m\over{4\pi {B}^2}}[{\r }^2-\vec \nabla \r
\cdot \int d{\vec {x^\prime}}\ \nabla  {1\over{\ve \vec x-\vec {x^\prime }
\ve }}\r(\vec {x^\prime })]\eqno(5.6).$$
This can be readily solved to obtain
$$a_0(\vec x,t)={1\over{4\pi }}\int d{\vec {x^\prime }}\ {1\over{\ve \vec x
-\vec {x^\prime }\ve }}\r_{\rm eff}(\vec {x^\prime },t)\eqno(5.7).$$
The terms in $\r_{\rm eff}$ beyond the first show the effects of
Landau level mixing.

Putting $a_0$ from (5.7) into (4.5), one obtains $J^i$ in the presence
of mixing.
$$\eqalignno{J^i(\vec x,t)&={1\over{4\pi  B}}\r(\vec x,t) \e^{0ij}\del_j
\bigl[\int d\vec{x^\prime }\ {1\over{\ve \vec x-\vec{x^\prime }\ve }}
\bigl( \r(\vec{x^\prime },t)-{m\over{{B}^2}}\bigl\{ \r^2(\vec
{x^\prime},t)\cr &-{\nabla }^\prime \r(\vec {x^\prime },t)\int d\vec y
({\nabla }^\prime
 {1\over{\ve \vec {x^\prime }-\vec y\ve }})\r(\vec y,t)
+ {1\over{4\pi }}(\int d{x^\prime }\ \nabla
{1\over{\ve \vec x-\vec {x^\prime }\ve }}\r(\vec {x^\prime },t))^2 \bigr\}
\bigr) \bigr] .&(5.8)\cr }$$
The first term agrees with that obtained in [4], as it should.
The terms beyond that are the corrections to $O(m/{\sqrt B})$, as mentioned in
[5].
\bigskip
\centerline{\bf VI. Minimal Coupling to Electromagnetism }
\bigskip
The effective action for planar electrons in a strong magnetic field
coupled minimally to perturbative slowly-varying electromagnetic
fields has already been given in [6],[7]. Here, we discuss the structure
of the current operators arising from this effective action as a
natural extension to what has been discussed in the earlier sections.
 From [6], we know that the effective action is given by
$$S_{\rm eff}=\int dt\ \int d\vec x\ \bigl[\bar \y(\vec x,t)i\del_t \y(\vec x,
t)-\r(\vec x,t)\bigl\{ a_0(\vec x,t)+{1\over B}\e_{0ij}(a^i\del^j a^0
-{1\over2}a^i\del^0 a^j)\bigr\} \bigr] \eqno(6.1).$$
If we also consider the Maxwell term governing $a^\m $, we can readily
obtain the Maxwell equations for the potentials. Here, we choose to work
in the $\nabla \vec A=0 $ gauge, where, $A\equiv {1\over{\sqrt{2B}}}(a^1
+ia^2)$ may be written as
$$\eqalignno{A&=-i\del_{\bar z}\f \cr
\bar A&=i\del_z \f .&(6.2)\cr }$$
The Maxwell equations are,
$$(\del_t^2-{\nabla }^2 +{{i}\over{2B}}\del_t \r )\del_{\bar z}\f
=-i\del_{\bar z}\del_t a_0 \eqno(6.3)$$
and
$${\nabla}^2 a_0=-\r +{1\over B}(\nabla \cdot \r \nabla \f +\r {\nabla }^2 \f )
\eqno(6.4).$$
We can solve these iteratively in $m/\sqrt{B} $. To O(1),
$$a_0\simeq {1\over{4\pi }}\int d\vec y\ {1\over{\ve \vec x-\vec y \ve }}
\r(\vec y,t)\eqno(6.5).$$
Further,
$$(\del_t^2-{\nabla }^2) \f =-{1\over{4\pi }}\int d\vec y\
{1\over{\ve \vec x-\vec y \ve }}\del_t \r(\vec y,t)\eqno(6.6).$$
We can solve (6.6) in momentum space. We get,
$$\eqalignno{a^1(\w , \vec k)&={{i\w k^2}\over{2\ve \vec k \ve ({\w }^2
+{\vec k}^2)}}\r(\w ,\vec k)\cr
a^2(\w ,\vec k )&={{-i\w k^1}\over{2\ve \vec k \ve ({\w }^2+{\vec k}^2)}}
\r(\w ,\vec k )&(6.7).\cr }$$
Putting (6.7) back in (6.4), we obtain
$$a_0(\w ,\vec k)={1\over{2\ve \vec k \ve }}\bigl[ \r(\w ,\vec k)
+{1\over{2B}}\int {{d\z d\vec p}\over{(2\pi )^{{3\over2}}}}\
{{\z \vec k\d. \vec p}\over{\ve \vec p \ve ({\z }^2+{\vec p}^2)}}
\r(\w -\z ,\vec k-\vec p,t)\r(\z ,\vec p,t)\bigr] \eqno(6.8). $$
The first term is the Coulomb term and the subsequent terms are corrections
due to Landau level mixing.
Now the appropriate current operator as obtained from (6.1) is
$$J=i\r \del_z [a_0+{1\over B}\e_{0ij}\{ a^i\del^j a^0-{1\over 2}
a^i\del_0a^j \} ]\eqno(6.9).$$
 From this it is quite clear that we need $a_0$ beyond the leading
order only in the first term in (6.9), and $a^i$ is never required
beyond what is given in (6.7). Thus using (6.7), (6.8) and (6.9),
we obtain the current operator in this case beyond the L.L.L.
approximation.
\bigskip
\centerline{\bf VII. Including a local four-fermi term }
\bigskip
A local four-fermi term is the ultralocal limit of a density-density
type interaction.
$$H_{\rm int}={g\over 2}\int d\vec x\ [\y(\vec x)^\dag \y(\vec x)]^2
\eqno(7.1).$$
In the previous two sections, we have utilised
the Maxwell equations to obtain corrections to a density-density
type of interaction even though the fermionic part of the theory
was bilinear in the fermionic fields. In this case, we may use
the extremely well-known auxilliary field method to render the
theory bilinear in the fermionic field.
Using an auxilliary field $\s(\vec x,t)$, we rewrite the above as
$$H_{\rm int}\rightarrow \int d\vec x\ [\y(\vec x)^\dag \y(\vec x)\s(\vec x)
-{1\over{2g}}\s^2(\vec x)]\eqno(7.2). $$
Thus as far as the fermionic integration is concerned, we are shifting
V in (3.9) to $\rm V+\s $. For simplicity, we set $\rm V\rightarrow 0 $
in the effective action. The Euler-Lagrange equation for $\s $ may
be solved iteratively as in the previous cases. To leading order,
$\s(\vec x,t)=g\r(\vec x,t)$, as anticipated. When this is re-inserted
in the equation of motion for $\s $, we obtain
$$\s =g[\r +{g\over{\w }}\{ \del_{\bar z}(\r \del_z \r )+\del_z (\r
\del_{\bar z}\r )\} ]\eqno(7.3).$$
This in turn, inserted into (4.5), yields the correction to the
current operator in the presence of the four-fermi term due to
mixing with higher Landau levels.
\bigskip
\centerline{\bf VIII. Edge Current in a Quantum Hall Droplet }
\bigskip
Given the expression for the current operator obtained earlier,
one could directly address the issue of the edge-current in a
finite quantum Hall droplet. The idea of using a confining
potential to create a finite droplet has already been elaborated
upon in [6]. Let us consider the potential to be of the form
$V=V_0 +U$, where $V_0=Ex .$ Further, imagine momentarily that
U has been set to zero. In that case, the effective Hamiltonian to
leading order is given by
$$H_{\rm eff}^{(0)}=E\int dX\ \y^\dag (X) X \y(X)
\eqno(8.1). $$
Here, we have used the basis where $\hat X $ is diagonal.
Further,
$$\{ \y(X), \y^\dag (X^\prime )\}=\d(X-X^\prime ) \eqno(8.2).$$
We define the droplet to be such that all the single-particle
states upto the zero-energy state are filled. Thus, $X=0$ is the Fermi
surface.
The corresponding many-body ground state is
$$\ve G \ke \equiv \prod_{X\leq0}\y^\dag (X)\ve 0 \ke \eqno(8.3).$$
As discussed in [6], the density operator, normal-ordered with respect
to this new ground state is given by
$$\r(z,\bar z,t)=\int_{-\infty}^\infty dX\int_{-\infty}^\infty
dX^\prime\ [:\y^\dag(X,t)\y(X^\prime,t):+
\Theta(-X)\d(X-X^\prime)]\br X\ve z\ke\br\bar z\ve X' \ke
e^{-\ve z\ve^2}\eqno(8.4)$$
The first term gives the contribution of the interior. The second term
is the edge contribution.
 From (4.5) and (8.4), we see that
$$J_{\rm edge}^y(x,y,t)\simeq -{E\over B}:\r(x,y,t): \eqno(8.5), $$
where $: \r :$ denotes the second term in (8.4). From (2.17) it is clear
that $: \r :$ is peaked sharply at $x=0$. Thus,
$$J_{\rm edge}^y(y,t)\simeq -{E\over B}: \y^\dag (y,t)\y(y,t) : \eqno(8.6)$$
to leading order, where
$$\{ \y(y), \y^\dag (y^\prime )\}=\d(y-y^\prime )\eqno(8.7).$$
This is just a chiral current moving with a velocity $v={E\over B}$ in
agreement with [7], [8].
\bigskip
\centerline{\bf IX. Conclusion }
\bigskip
In this paper, we have studied how the interactions of fermions in the
L.L.L. get modified due to mixing with higher Landau levels. Specifically,
we have concentrated on the current operator and have deduced its form for
a variety of interactions, including the long-range Coulomb interaction,
electromagnetic interactions and contact interaction between the fermions.
We have given a method of systematically including corrections arising
from Landau level mixing. For electromagnetic interactions, these corrections
introduce $1/r^2$, $1/r^3$ and higher modifications to the fundamental $1/r$
interaction and this is reflected in the current and density profiles of
fermions, e.g. in the quantum Hall effect [6].
It is interesting to note that Landau
level mixing introduces three-body and higher interactions between the fermions
when the interactions are at most two-body in the absence of mixing.
In conclusion we should also mention that the expression for the L.L.L.
current has enabled us to extract the form of the edge-current operator
in the case of a Hall sample of finite geometry.
\bigskip
\centerline{\bf Acknowledgements }
R.R. wishes to express his sincere appreciation of the discussions he
had on this subject with V.P. Nair. He also thanks B. Sakita for introducing
him to the exciting topic of L.L.L. fermions. G.G. and R.R. thank the
National Science Foundation for partial support while this work was
in progress.
\vfill\eject
\centerline{\bf References }
\bigskip
\item{[1]} M. Stone,\lq\lq Quantum Hall Effect'' (World Scientific, Singapore,
1992).
\item{[2]} S. Iso, D. Karabali \& B. Sakita, Nucl. Phys.{\bf B 388},700,
(1992); Phys. Lett. {\bf B296},143,(1992).
\item{[3]} S.M. Girvin \&  T. Jach, Phys. Rev. {\bf B 29},5617,(1983).
\item{[4]} J. Martinez \& M. Stone, NSF-ITP 93-38.
\item{[5]} R. Rajaraman, \lq\lq Currents in the Lowest Landau Level Field
Theory with e-e Interactions,"  Urbana-Champaign preprint (1993).
\item{[6]} S.L. Sondhi \& S.A. Kivelson, Phys. Rev. {\bf B 46},13319,
(1992).
\item{[7]} R. Ray \& B. Sakita, \lq\lq Electromagnetic Interactions of
electrons in the lowest Landau level," hep-th 9304033, (to be published),
{\it Ann. of Physics}.
\item{[8]} B. Sakita, \lq\lq  $W_{\infty }$ gauge transformations and the
electromagnetic interactions of electrons in the lowest Landau level,"
CCNY-HEP-93/2, hep-th 9307087.

\end